\documentclass[english,pra,showpacs,twocolumn,amsmath]{revtex4}
\usepackage{graphicx}
\usepackage{simplewick}
\usepackage{latexsym}

\begin{document}
\title{Lambda transition and Bose-Einstein condensation in liquid ${\rm ^4He}$}

\author{Vladimir I. Kruglov}

\affiliation{Centre for Engineering Quantum Systems, 
School of Mathematics and Physics, The University of Queensland, Brisbane, QLD 4072, Australia}

\begin{abstract}
We present a theory describing the lambda transition and the Bose-Einstein condensation (BEC) in a liquid ${\rm ^4He}$ based on the diatomic quasiparticle concept. It is shown that in liquid ${\rm ^4He}$ for the temperature region $1~{\rm K}\leq{\rm T}\leq {\rm T}_\lambda$ the diatomic quasiparticles macroscopically populate the ground state which leads to BEC in liquid ${\rm ^4He}$. The approach yields the lambda transition temperature as ${\rm T}_\lambda=2.16~{\rm K}$ which is in excellent agreement with the experimental lambda temperature ${\rm T}_\lambda=2.17~{\rm K}$. The concept of diatomic quasiparticles also leads to superfluid and BEC fractions which are in a good agreement with the experimental data and Monte Carlo simulations.

\end{abstract}


\maketitle

\section{Introduction}

The similarity between liquid ${\rm ^4He}$ and an ideal Bose-Einstein gas was recognised by London. He assumed that the lambda transition in liquid ${\rm ^4He}$ is the analog for a phase transition in an ideal Bose gas at low temperature \cite{Lon,Lond}. This idea supports by estimation based on an equation for critical temperature in an ideal Bose gas.
Thereafter Tisza suggested that the presence of the condensed particles can be described by a two-fluid hydrodynamics \cite{Tis}. In this model the ``condensate" completely has no friction, while the rest behave like an ordinary liquid. However, Tisza's model did not appear to be completely self consistent and quantitative. 

Two-fluid quantitative hydrodynamics was subsequently developed by Landau \cite{Lan}. However in this paper is not assumed the idea of BEC (Bose-Einstein condensation). Landau also has predicted the excitation spectrum of liquid He II which goes over from the phonon behaviour $\epsilon(p)=cp$ at small momenta to a ``roton-like" form at larger values of the momenta as $\epsilon(p)=\Delta+(p-p_0)^2/2\mu_R$.
This phenomenological model was based mostly on experimental data and deep physical intuition. Modern understanding of superfluidity is based on the Onsager-Feynman quantisation condition which is also important for the two-fluid theory in liquid ${\rm ^4He}$  \cite{Feyn}. 
The excitation spectrum in liquid ${\rm ^4He}$ was measured in neutron scattering experiments with great accuracy by several groups, in particular by Henshaw and Woods \cite{Hen}. This spectrum qualitatively agrees with Landau's phenomenological excitation spectrum.
 
 The Bogoliubov analytical results \cite{Bog} for elementary excitations in the Bose gas are very important for understanding the excitation spectrum in phonon region. Nevertheless the Bogoliubov's theory can be applied only to dilute Bose systems. 
Feynman has found a relation between the energy spectrum of the elementary excitations and the structure factor \cite{Fe,Fey} that verifies Landau's phenomenological dispersion relation. We note that the Feynman's relation is correct only for small enough momenta when the excitations are phonons. In a ``roton-like" region it is only an approximation of a real situation. 
Feynman has also proposed a model of the roton excitation as localised vortex ring \cite{Feyn} with a characteristic size of the order of the mean atomic distance in liquid ${\rm ^4He}$. A number of methods have been also suggested for applications to quantum Bose liquids. For a review of quantum fluid theories also see the references \cite{Gav,D,N,G,A,B,Ba,L,Kru,Krug,Gi, Kal}.

The use of neutrons to observe the condensate fraction in liquid ${\rm ^4He}$ was proposed in \cite{Mil,Hoh} and
the first measurement of BEC fraction in liquid ${\rm ^4He}$ has been reported by Cowley and Woods \cite{Cow}. The history of measurement to higher incidence energy neutrons and improved spectrometer performance is reviewed by Glyde \cite{Gly}, Sokol \cite{Sok} and others. The recent measurements of BEC and the atomic momentum distribution in liquid and solid ${\rm ^4He}$ is reported by Diallo, Glyde and others \cite{D1,D2,D3,D4}.
The momentum distribution of liquid ${\rm ^4He}$ and ${\rm ^3He}$ has been calculated, at zero temperature, by using the Green-function Monte Carlo method \cite{Whi} and the diffusion Monte Carlo \cite{Mor, Moro}. At finite temperature, the simulations have been done by the path integral Monte Carlo method \cite{Cep, Poll}. For bosons they provide energy estimates that are virtually exact, within statistical accuracy. The optimisation procedure based on Monte Carlo calculations has been proposed in \cite{Viti}. 

While the above mentioned results provide the conceptual basis for understanding of superfluidity and BEC in liquid ${\rm ^4He}$ there are  fundamental questions are still open in this area. For example, it is often assumed that the lambda transition in liquid ${\rm ^4He}$ is the analog for a phase transition in an ideal Bose gas. However it was never been proven as a consequence of the Bose statistics of strong interacting atoms in  liquid ${\rm ^4He}$. 
Moreover there is no analytic quantitative theory describing the lambda transition phenomenon and the Bose-Einstein condensation in liquid ${\rm ^4He}$ around the lambda transition temperature. 
 
In this paper we develop the theory for superfluid ${\rm ^4He}$ based on a diatomic quasiparticles (DQ) concept. The DQ's coupling in liquid ${\rm ^4He}$ for the temperatures below lambda transition can be explained by the Lennard-Jones intermolecular potential. There is no spin interaction in this case because ${\rm ^4He}$ atoms are bosons with zero spin. We note that the pairs of coupled quasiparticles are observed in superfluid ${\rm ^3He}$ with spin $S=1$ and orbital momentum $L=1$. 
In this case the spin interaction yields the coupling in superfluid ${\rm ^3He}$ which is also known as spin triplet or odd parity pairing \cite{An,And,Bal,EnH}.
The approach based on diatomic quasiparticles concept yields the critical BEC temperature which is very close to experimental lambda transition temperature in liquid ${\rm ^4He}$. This theory also leads to superfluid and BEC fractions for liquid ${\rm ^4He}$ which are in a good agreement with recent condensation measurements \cite{D1,D2,D3,D4,Gl} and Monte Carlo simulations for a wide range of temperatures. 

In Section II we develop the theory for coupled quasiparticles in liquid ${\rm ^4He}$ for the temperatures below lambda transition. This many-body approach yields the discrete energy spectrum and the effective mass $M_q$ for DQ which are the bound states of two helium atoms interacting with the atoms of the bulk. We also present in this section the ground state energy of DQ. This energy is connected to the roton gap $\Delta$ as $E_0=-\Delta$.
In Section III we derive the thermodynamical functions and the equation for critical temperature of BEC in liquid ${\rm ^4He}$. This equation yields ${\rm T}_c=2.16~{\rm K}$ which is very close to the experimental value of lambda transition temperature ${\rm T}_\lambda=2.17~{\rm K}$.
In Section IV we present the theory of BEC  in liquid ${\rm ^4He}$ based on the DQ concept. It is shown that the DQ's condense at negative ground state energy $E_0$. We also calculate in this section the excited and condensed densities of diatomic quasiparticles and the energy and entropy of DQ's for the temperature region ${\rm T}_p\leq{\rm T}\leq {\rm T}_\lambda$ with ${\rm T}_p\simeq 1~{\rm K}$.
In Sections V and VI the superfluid and BEC fractions are calculated for temperature regions ${\rm T}_p \leq {\rm T} \leq {\rm T}_\lambda$ and ${\rm T} \leq 0.5~{\rm K}$ respectively. We show in these sections that the theoretical superfluid and BEC fractions are in a good agreement with experimental data and Monte Carlo simulations \cite{Whi,Mor,Moro,Cep,Poll,Mas}.

\section{Diatomic quasiparticles}

We describe in this section the spectrum of diatomic quasiparticles in liquid ${\rm ^4He}$ for the temperatures ${\rm T}_p\leq{\rm T}\leq {\rm T}_\lambda$ with ${\rm T}_p\simeq 1~{\rm K}$. The diatomic quasiparticles are defined as the bound states of two helium atoms interacting with the particles of the bulk. The DQ concept assumes that the number of diatomic quasiparticles is much less than the number of real particles.

The Hamiltonian for a many-body Bose system with two-particle potential $U(|{\bf r}_{i}-{\bf r}_{j}|)$ is of the form,
\begin{equation}
{\rm H}_N=\sum_{i=1}^{N}\frac{{\bf p}_i^{2}}{2m}+\sum_{i<j}^{N}U(|{\bf r}_{ij}|),
\label{1}
\end{equation}
where ${\bf r}_{ij}={\bf r}_{i}-{\bf r}_{j}$. We assume the commutator relation $[r_{s\alpha},p_{n\beta}]=i\hbar\delta_{sn}\delta_{\alpha\beta}$, and all other commutators are zero. Here $s,n$ and $\alpha,\beta$ are the the particle numbers and the projector indexes respectively. 

The Hamiltonian in Eq. (\ref{1}) can also be written as
\begin{equation}
{\rm H}_N={\rm H}_q+{\rm H}_{N-2}, 
\label{2}
\end{equation}
where ${\rm H}_q$ is the Hamiltonian for two bound particles with the numbers $1$ and $2$  interacting with all particles of the bulk, and ${\rm H}_{N-2}$ is the Hamiltonian for the rest particles of the many-body Bose system. It follows from Eq. (\ref{1}) that the Hamiltonians ${\rm H}_q$ and 
${\rm H}_{N-2}$ are
\begin{equation}
{\rm H}_q=\frac{{\bf p}_1^{2}}{2m}+\frac{{\bf p}_2^{2}}{2m}+U(|{\bf r}_{12}|)+\sum_{j=3}^{N}U(|{\bf r}_{1j}|)+\sum_{j=3}^{N}U(|{\bf r}_{2j}|),
\label{3}
\end{equation}
\begin{equation}
{\rm H}_{N-2}=\sum_{i=3}^{N}\frac{{\bf p}_i^{2}}{2m}+\sum_{3\leq i<j}^{N}U(|{\bf r}_{ij}|).
\label{4}
\end{equation}

We define the canonical transformation for the operators ${\bf r}_s$ and ${\bf p}_s$  (s=1,2) as 
\begin{equation}
{\bf R}= \frac{1}{2}({\bf r}_1+{\bf r}_2),~~ {\bf r}={\bf r}_1-{\bf r}_2,
\label{5}
\end{equation}
\begin{equation}
{\bf P}= {\bf p}_1+{\bf p}_2,~~ {\bf p}=\frac{1}{2}({\bf p}_1-{\bf p}_2).
\label{6}
\end{equation}
The commutation relations  for these operators are $[r_\alpha,p_\beta]=i\hbar\delta_{\alpha\beta}$ and $[R_\alpha,P_\beta]=i\hbar\delta_{\alpha\beta}$, and all other commutators are zero. 
The Hamiltonian given by Eq. (\ref{3}) can be written as
\begin{equation}
{\rm H}_q=\frac{1}{2m_c}{\bf p}^{2}+\frac{1}{2M}{\bf P}^{2}+{\cal U}({\bf R},{\bf r},{\bf X}_{N-2}),
\label{7}
\end{equation}
where $m_c=m/2$ and $M=2m$. The potential ${\cal U}({\bf R},{\bf r},{\bf X}_{N-2})$ in this equation is of the form,
\begin{eqnarray}
{\cal U}({\bf R},{\bf r},{\bf X}_{N-2})=U(|{\bf r}|)+\sum_{j=3}^{N}U(|{\bf R}+\frac{1}{2}{\bf r}-{\bf r}_j|)~~~
\nonumber\\ \noalign{\vskip3pt}+\sum_{j=3}^{N}U(|{\bf R}-\frac{1}{2}{\bf r}-{\bf r}_j|),~~~~~~~~~~
\label{8}
\end{eqnarray}
with ${\bf X}_{N-2}=({\bf r}_3,{\bf r}_4,...,{\bf r}_N)$. 

The intermolecular interaction for Bose fluid is given by Lennard-Jones potential,
\begin{equation}
U(r)=4\epsilon\left[\left(\frac{r_0}{r}\right)^{12}-\left(\frac{r_0}{r}\right)^{6}\right],
\label{9}
\end{equation}
where the minimum of the potential occurs at $r_m=2^{1/6}r_0$. In the case of a gas or liquid ${\rm ^4He}$ to good accuracy, the parameters of the Lennard-Jones potential calculated by a self-consistent-field Hartree-Fock method
are given by $\epsilon/k_B=10.6~{\rm K}$ and $r_m=2.98\cdot 10^{-8}~{\rm cm}$ \cite{Ah,Az,Azi}.

We use below in this section the Schr\"odinger representation for the operators. The potential in Eq. (\ref{8}) can be written,
\begin{equation}
{\cal U}({\bf R},{\bf r},{\bf X}_{N-2})=\bar{{\cal U}}({\bf R},{\bf r})+{\cal V}({\bf R},{\bf r},{\bf X}_{N-2}),
\label{10}
\end{equation}
where $\bar{{\cal U}}({\bf R},{\bf r})=\langle{\cal U}({\bf R},{\bf r},{\bf X}_{N-2})\rangle_{N-2}$ is the potential 
averaging by the density matrix over the position of $N-2$ particles with the numbers $3,4,,..,N$.
The function ${\cal V}({\bf R}\,{\bf r},{\bf X}_{N-2})$ describes the fluctuations of the potential in Eq. (\ref{10}) with 
$\langle{\cal V}({\bf R},{\bf r},{\bf X}_{N-2})\rangle_{N-2}=0$. 

The decomposition of the average potential $\bar{{\cal U}}({\bf R},{\bf r})$ of the quasiparticle in a series around an equilibrium position with ${\bf R}={\bf R}_0$ and $r=r_0$ has the form,
\begin{eqnarray}
\bar{{\cal U}}({\bf R},{\bf r})=\bar{\cal U}_0+\frac{1}{2}m_c\omega_0^2(r-r_0)^2 + 
\nonumber\\ \noalign{\vskip3pt}     +\sum_{s=1}^{3} \frac{1}{2}M_q\omega_s^2 (R_s-R_{0s})^2+ ...~,
\label{11}
\end{eqnarray}
where $\bar{\cal U}_0=\bar{{\cal U}}({\bf R}_0,{\bf n}r_0)$ and ${\bf n}={\bf r}/r$ is the unit vector. The frequencies are  given by $\omega_0^2=m_c^{-1}\partial_r^2\bar{{\cal U}}({\bf R}_0,{\bf n}r)|_{r=r_0}$ and $\omega_s^2=M_q^{-1}\partial_{R_s}^2\bar{{\cal U}}({\bf R},{\bf n}r_0)|_{{\bf R}={\bf R}_0}$ for $s=1,2,3$ respectively.
We use  in this decomposition the renormalised mass $M_q=\sigma M$ for diatomic quasiparticles. 
The factor is $\sigma=2^{2/3}$ (see Appendix A) and hence the renormalised mass is given by $M_q=\sigma M=2^{5/3}m$.
We also note that the frequency $\omega_s$ depends on the temperature and the density of the Bose liquid.

The  DQ Hamiltonian can be written by Eqs. (\ref{7}), (\ref{10}) and (\ref{11}) in the form,
\begin{equation}
{\rm H}_q=\bar{\cal U}_0+{\rm H}_q^{(1)}+{\rm H}_q^{(2)}+{\cal V}({\bf R},{\bf r},{\bf X}_{N-2}),
\label{12}
\end{equation}
where the Hamiltonians ${\rm H}_q^{(1)}$ and ${\rm H}_q^{(2)}$ are

\begin{eqnarray}
{\rm H}_q^{(1)}=-\frac{\hbar^2}{2m_c}\frac{1}{r^2}\frac{\partial}{\partial r}\left( r^2 \frac{\partial}{\partial r} \right)+\frac{1}{2}m_c\omega_0^2(r-r_0)^2
\nonumber\\ \noalign{\vskip3pt} +\frac{\hbar^2}{2m_cr_0^2}{\rm J}({\rm J}+1),~~~~~~~~~~
\label{13}
\end{eqnarray}
\begin{equation}
{\rm H}_q^{(2)}=-\frac{\hbar^2}{2M_q}\sum_{s=1}^{3}\frac{\partial^2}{\partial R_s^2}+\sum_{s=1}^{3}\frac{1}{2}M_q\omega_s^2(R_s-R_{0s})^2.
\label{14}
\end{equation}
The Hamiltonian in Eq. (\ref{13}) describing the internal degrees of freedom of DQ is written in spherical coordinate system. We note that the
term ${\cal V}({\bf R},{\bf r},{\bf X}_{N-2})$ is connected with the fluctuation of interactions in Eq. (\ref{12}), and it can be neglected for the eigenenergy problem to a good accuracy. 
The effective Hamiltonian ${\rm H}_q^{(1)}$
describes the internal $1{\rm D}$ vibrations and rotations of DQ and the Hamiltonian ${\rm H}_q^{(2)}$ describes $3{\rm D}$ harmonic vibrations for the centre of mass of DQ with the effective mass $M_q$. 

The discrete energy spectrums follow from the Hamiltonians ${\rm H}_q^{(1)}$ and ${\rm H}_q^{(2)}$ as $E_{n_0{\rm J}}^{(1)}=(\hbar^2/2m_cr_0^2){\rm J}({\rm J}+1)+\hbar\omega_0(n_{0}+1/2)$ and   
$E_{n_1n_2n_3}^{(2)}=\sum_{s=1}^3\hbar\omega_s (n_s+1/2)$ respectively where ${\rm J}=0,1,2,...$ and $n_s=0,1,2,...$ with $s=0,1,2,3$. 
However the rotations of DQ are 'frozen' in the liquid at low temperatures. Hence the angular momentum quantum number is ${\rm J}=0$. Moreover one can choose the third cartesian axis as ${\bf e}_3={\bf n}$, then the average potential $\bar{{\cal U}}({\bf R},{\bf n}r)$ yields the relation $\omega_1=\omega_2$ for the frequencies. 
The discrete eigenenergies of the Hamiltonian ${\rm H}_q$ are 
\begin{equation}
E_{n_0n_1n_2n_3}=\bar{\cal U}_0+E_{n_0{\rm 0}}^{(1)}+E_{n_1n_2n_3}^{(2)}=E_{0}+\sum_{s=0}^{3}\hbar\omega_sn_s,
\label{15}
\end{equation}
where $E_0=\bar{\cal U}_0+\sum_{s=0}^{3} \frac{1}{2}\hbar\omega_s$ is the energy of the ground state of DQ. 
We note that the DQ states with the energies $E_{n_0n_1n_2n_3}$ are metastable because the fluctuation of interaction given by the term ${\cal V}({\bf R},{\bf r},{\bf X}_{N-2})$ leads to collisions of a pair of bound atoms with surrounding atoms.
There is also another scattering channel which correspond to dissociation of DQ's.
The energy spectrum of the Hamiltonian ${\rm H}_q$ also has the continuous component given by
\begin{equation}
\varepsilon_q({\bf p})=\epsilon_q+\frac{{\bf p}^2}{2M_q},
\label{16}
\end{equation}
where ${\bf p}$ is the momentum of DQ. 
 
The Eq. (\ref{8}) yields the average potential energy in the form $\bar{{\cal U}}({\bf R},{\bf r})=U(r)+W({\bf R},{\bf r})$ where the function $W({\bf R},{\bf r})$ is given by 
\begin{equation}
W({\bf R},{\bf r})=\langle [U(|{\bf R}+\frac{1}{2}{\bf r}-{\bf r}'|) + U(|{\bf R}-\frac{1}{2}{\bf r}-{\bf r}'|)] \rangle_{{\bf r}'}.
\label{17}
\end{equation}
Here $\langle...\rangle_{{\bf r}'}$ is the averaging to variable ${\bf r}'$ with a conditional distribution. The position of the atoms in DQ are given by ${\bf r}_1={\bf R}+\frac{1}{2}{\bf r}$, and ${\bf r}_2={\bf R}-\frac{1}{2}{\bf r}$. It follows from Eq. (\ref{17})
that the function $W({\bf R},{\bf r})$ is  invariant to the change ${\bf r}_1\rightarrow {\bf r}_2$ and ${\bf r}_2\rightarrow {\bf r}_1$. 

The inequality $\hbar\omega_s\ll |E_0|$ (see Appendix B) yields the relation $W({\bf R},{\bf n}r_m)\simeq W({\bf R},{\bf n}r_a)$ where $r_m=2^{1/6}r_0$, and $r_a=2a_0$ at low pressures \cite{Krug}. Here $a_0$ is the s-scattering length of helium atoms. Thus the ground state energy of DQ is
 \begin{eqnarray}
E_0=\bar{{\cal U}}({\bf R},{\bf n}r_m)-\bar{{\cal U}}({\bf R},{\bf n}r_a)~~~~~~~~~~~~~~
\nonumber\\ \noalign{\vskip3pt} \simeq U(r_m)-U(r_a)= -4\epsilon\left[\frac{1}{4}-\left(\frac{r_0}{r_a}\right)^{6}+\left(\frac{r_0}{r_a}\right)^{12}\right].
\label{18}
\end{eqnarray}

The relation $r_a=2a_0$ is connected to s-scattering cross-section $\sigma_s=4\pi a_0^2$ for helium atoms. We note that the s-scattering length for ${\rm ^4He}$ atoms is $a_0=2.2\cdot 10^{-8}~{\rm cm}$ \cite{Kr}. Moreover the distance $r_a$ is close to the average distance $\bar{r}=2(3m/4\pi\rho)^{1/3}$ between the atoms in the Bose liquid at low pressure. 
Hence for enough low pressures we have the relation $r_a=a\bar{r}$. This equation leads to the ground state energy $E_0$ of DQ's as a function of the mass density $\rho$.
The constant parameter $a$ can be defined as $a=2a_0/\bar{r}$, where the average distance $\bar{r}$ is given for the mass density $\rho=0.145$ ${\rm g~cm^{-3}}$. In this case the average distance is $\bar{r}=4.44\cdot 10^{-8}~{\rm cm}$, and hence the constant parameter is $a=0.991$.

The Eq. (\ref{18}) leads to the ground state energy of DQ's which is a function of mass density as
\begin{equation}
E_0(\rho)=-\epsilon\left[1-4\left(\frac{\rho}{\rho_a}\right)^{2}+4\left(\frac{\rho}{\rho_a}\right)^{4}\right],~~~\rho_a=\frac{6ma^3}{\pi r_0^3}.
\label{19}
\end{equation}
We note that for low pressure the mass density is $\rho=0.145$ ${\rm g~cm^{-3}}$, and hence the Eq. (\ref{19}) yields $E_0/k_B=-8.65~{\rm K}$. Moreover the roton gap $\Delta$ in liquid ${\rm ^4He}$ is connected to the ground state energy of DQ's by relation $\Delta=-E_0$ \cite{Krug}.

\section{Lambda transition temperature in liquid ${\rm ^4He}$}

In this section we derive the equations for the average energy, free energy and entropy for the trapped DQ in liquid ${\rm ^4He}$. We also derive the equation for the lambda transition temperature which agrees to a high accuracy with experimental temperature ${\rm T}_\lambda$ for liquid ${\rm ^4He}$. This equation is found as a necessary condition for the lambda transition. It is shown in the next section that this condition is sufficient as well.

We also show in Section V that the mass density of DQ's is much smaller than the mass density $\rho$ in liquid ${\rm ^4He}$. In this case the probability that trapped DQ has the vibrational energy $E_{n_0n_1n_2n_3}$ is given by
\begin{equation}
{\cal P}_{n_0n_1n_2n_3}(\beta)=\frac{1}{{\cal Z}(\beta)}\exp(-\beta E_{n_0n_1n_2n_3}),
\label{20}
\end{equation}
where $\beta=1/k_B {\rm T}$ and the partition function ${\cal Z}(\beta)$ is
\begin{equation}
{\cal Z}(\beta)=\sum_{n_0=0}^\infty\sum_{n_1=0}^\infty\sum_{n_2=0}^\infty\sum_{n_3=0}^\infty\exp(-\beta E_{n_0n_1n_2n_3}).
\label{21}
\end{equation}
We note that the discrete energy spectrum in Eq. (\ref{15}) is accurate only for enough small quantum numbers $n_s$. However the partition function in Eq. (\ref{21}) is given to good accuracy because the terms with large numbers $n_s$ are small when 
${\rm T}\leq {\rm T}_{\lambda}$. The calculation of the sums in Eq. (\ref{21}) leads to the equation for partition function ${\cal Z}(\beta)$ as
\begin{equation}
{\rm ln}{\cal Z}(\beta)=-\beta E_0-\sum_{s=0}^3 {\rm ln}[1-  \exp(-\beta \varepsilon_s)],
\label{22}
\end{equation}
with $\varepsilon_s=\hbar\omega_s$. Thus the average energy of DQ and the dispersion of the energy are
\begin{equation}
\bar{E}_q(\beta) =-\frac{\partial}{\partial\beta} {\rm ln}{\cal Z}(\beta)=E_0+\sum_{s=0}^3 \frac{\varepsilon_s}{\exp(\beta \varepsilon_s)-1},
\label{23}
\end{equation}
\begin{equation}
D_q(\beta)=\frac{\partial^2}{\partial\beta^2} {\rm ln}{\cal Z}(\beta)=\sum_{s=0}^3 \frac{\varepsilon_s^2\exp(\beta \varepsilon_s)}{[\exp(\beta \varepsilon_s)-1]^2}.
\label{24}
\end{equation}
The free energy and the entropy of DQ are given by 
\begin{equation}
F_q(\beta)=-\frac{1}{\beta}{\rm ln}{\cal Z}(\beta)=E_0+\frac{1}{\beta}\sum_{s=0}^3 {\rm ln}[1-  \exp(-\beta \varepsilon_s)],
\label{25}
\end{equation}

\begin{equation}
S_q(\beta)= -\frac{\partial}{\partial {\rm T}} F_q(\beta)={\rm T}^{-1}(\bar{E}_q(\beta)- F_q(\beta)).
\label{26}
\end{equation}

These equations can be simplified for the temperatures ${\rm T}\simeq {\rm T}_{\lambda}$. We show in the Appendix B that the next inequality is satisfied for liquid ${\rm ^4He}$,
\begin{equation}
\beta_\lambda \varepsilon_s=\frac{ \hbar\omega_s}{k_B{\rm T}_\lambda}\ll 1,~~~s=0,1,2,3.
\label{27}
\end{equation}
Thus for temperatures ${\rm T}\simeq {\rm T}_{\lambda}$ the Eq. (\ref{23}) and (\ref{24}) can be written as
\begin{equation}
\bar{E}_q(\beta) =E_0+4k_B{\rm T},
\label{28}
\end{equation}
\begin{equation}
\Delta_q(\beta)=\sqrt{ D_q(\beta)}=2k_B{\rm T},
\label{29}
\end{equation}
where $\Delta_q(\beta)$ is the energy variation. In the case ${\rm T}\simeq {\rm T}_{\lambda}$ the free energy and entropy of DQ are given by
\begin{equation}
F_q(\beta)=E_0+k_B{\rm T}\sum_{s=0}^3 {\rm ln}\frac{\varepsilon_s}{k_B{\rm T}},
\label{30}
\end{equation}
\begin{equation}
S_q(\beta)=4k_B-k_B\sum_{s=0}^3 {\rm ln}\frac{\varepsilon_s}{k_B{\rm T}}.
\label{31}
\end{equation}
These equations lead to the entropy differential as
\begin{equation}
dS_q(\beta)=\frac{4k_Bd{\rm T}}{{\rm T}}=\frac{d\bar{E}_q(\beta)}{{\rm T}}.
\label{32}
\end{equation}

The necessary condition for existing a finite fraction of trapped DQ's in liquid ${\rm ^4He}$ is $\bar{E}_q(\beta)<0$ which can be written by Eq. (\ref{28}) as $E_0+4k_B{\rm T}<0$. 
Thus the trapped DQ's have the finite fraction in liquid ${\rm ^4He}$ when the condition ${\rm T}<|E_0|/4k_B$ is satisfied. We show in the next section that this condition is also the sufficient condition for lambda transition in liquid ${\rm ^4He}$. 
Hence the critical temperature for BEC in liquid ${\rm ^4He}$ is given by
\begin{equation}
{\rm T}_c=\frac{|E_0(\rho)|}{4k_B},
\label{33}
\end{equation}
where the energy $|E_0|$ is defined by Eq. (\ref{19}) as
\begin{equation}
|E_0(\rho)|=\epsilon\left[1-4\left(\frac{\rho}{\rho_a}\right)^{2}+4\left(\frac{\rho}{\rho_a}\right)^{4}\right].
\label{34}
\end{equation}

It is accepted  that the critical temperature for BEC in liquid ${\rm ^4He}$ is equal to lambda transition temperature \cite{N}. 
The Eq. (\ref{34}) yields $|E_0|/k_B=8.65~{\rm K}$ for low pressure or the helium mass density $\rho=0.145$ ${\rm g~cm^{-3}}$. Hence in this case the Eq. (\ref{33}) leads to the critical BEC temperature as ${\rm T}_c={\rm T}_\lambda=2.16~{\rm K}$ which is in excellent agreement with the experimental lambda transition temperature given by ${\rm T}_\lambda=2.17~{\rm K}$.

The lambda transition temperature and the critical BEC temperature in Eqs. (\ref{33}) and (\ref{34}) are the functions of the mass density as  
\begin{equation}
{\rm T}_\lambda(\rho)={\rm T}_c(\rho)=\frac{\epsilon}{k_B}\left[\frac{1}{4}-\left(\frac{\rho}{\rho_a}\right)^{2}+\left(\frac{\rho}{\rho_a}\right)^{4}\right].
\label{35}
\end{equation}
where $\rho_a=6ma^3/\pi r_0^3$.
We emphasise that the theory does not applied to dilute Bose gas because the DQ's exist only
in liquid ${\rm ^4He}$ at low temperatures.

\section{BEC condensation in Bose fluids}

In this section we develop the theory of BEC in liquid ${\rm ^4He}$ for the temperature region ${\rm T}_p \leq {\rm T} \leq {\rm T}_\lambda$ where the bound temperature is ${\rm T}_p\simeq 1~{\rm K}$. The DQ's in this temperature region can be described as the Bose system consisting of two fractions. The first fraction is the Bose gas of DQ's with the continuos energy spectrum given by Eq. (\ref{16}) and the second fraction consists of the trapped DQ's with a discrete energy spectrum given in Eq. (\ref{15}).

The full Hamiltonian for  liquid Bose fluid can be written in the form,
\begin{equation}
\hat{{\rm H}}=\hat{{\rm H}}_{q}+\hat{{\rm H}}_a+\hat{{\rm H}}_{\rm int},
\label{36}
\end{equation}
where $\hat{{\rm H}}_{q}$ is the Hamiltonian describing DQ's in liquid ${\rm ^4He}$, and $\hat{{\rm H}}_a$ is the Hamiltonian for the rest free quasiparticles including the rotons and phonons. The Hamiltonian $\hat{{\rm H}}_{\rm int}$ describes the interaction of all sorts of quasiparticles in the Bose fluid. 

The concept of the quasiparticles assumes that the number of quasiparticle excitations is much less than the number of real particles. Thus the DQ's subsystem can be described as two fractions: an ideal Bose gas of DQ's with the continuos energy spectrum and the fraction of DQ's with a discrete energy spectrum. We emphasize that for this quasiparticles representation
the interaction energy of DQ's is much less than their full energy. Thus the DQ's are weakly interacting excitations in liquid Bose fluid. We show in Section V that this picture leads to good agreement with experimental data.

Using the results of Section II we can write the Hamiltonian for two components of DQ's as
\begin{eqnarray}
\hat{{\rm H}}_{q}=\sum_{{\bf p}}\varepsilon_q({\bf p})\hat{d}^{\dag}({\bf p})\hat{d}({\bf p})~~~~\nonumber\\ \noalign{\vskip3pt}+\sum_{n_0}\sum_{n_1}\sum_{n_2} \sum_{n_3}\delta_{\{n\}}E_{\{n\}}\hat{d}^{\dag}_{\{n\}}\hat{d}_{\{n\}},
\label{37}
\end{eqnarray}
where $E_{\{n\}}=E_{n_0n_1n_2n_3}$ is the discrete energy spectrum given by Eq. (\ref{15}) and $\varepsilon_q({\bf p})=\epsilon_q+{\bf p}^2/2M_q$ is the continuos energy spectrum of DQ's with $|\epsilon_q|\ll |E_0|$. The operators $\hat{d}^{\dag}_{\{n\}},~\hat{d}_{\{n\}}$ and $\hat{d}^{\dag}({\bf p}),~\hat{d}({\bf p})$ are creation and annihilation Bose operators for discrete and continuos energy spectrums of DQ's respectively. The function $\delta_{\{n\}}$ is defined as:
 $\delta_{\{n\}}=1$ when $E_{\{n\}}<0$ and $\delta_{\{n\}}=0$ otherwise. 
The number operators of DQ's for discrete and continuos energy spectrums have the form,  
\begin{equation}
\hat{{\rm N}}_{0}=\sum_{n_0}\sum_{n_1}\sum_{n_2} \sum_{n_3}\delta_{\{n\}}\hat{d}^{\dag}_{\{n\}}\hat{d}_{\{n\}},
\label{38}
\end{equation}
\begin{equation}
\hat{{\rm N}}_{1}=\sum_{{\bf p}}\hat{d}^{\dag}({\bf p})\hat{d}({\bf p})=\sum_{{\bf p}}\hat{{\rm N}}_{1}({\bf p}),
\label{39}
\end{equation}
where $\hat{{\rm N}}_{1}({\bf p})=\hat{d}^{\dag}({\bf p})\hat{d}({\bf p})$ and $\hat{{\rm N}}_{q}=\hat{{\rm N}}_{0}+\hat{{\rm N}}_{1}$ is the full number operator for the DQ's subsystem. 

The grand canonical density operator for the DQ's subsystem is
\begin{equation}
\hat{\rho}_q=\Xi_q^{-1}\exp[-\beta(\hat{{\rm H}}_{q}-\mu_{0}\hat{{\rm N}}_{0}-\mu_1\hat{{\rm N}}_1)],
\label{40}
\end{equation}
where the grand canonical partition function is
\begin{equation}
\Xi_q={\rm Tr}\exp[-\beta(\hat{{\rm H}}_{q}-\mu_{0}\hat{{\rm N}}_{0}-\mu_1\hat{{\rm N}}_1)].
\label{41}
\end{equation}
The chemical potentials $\mu_0$ and $\mu_1$ are equal $\mu_0=\mu_1=\tilde{\mu}$. Hence 
the average number $\tilde{{\rm N}}_{1}({\bf p})=\langle \hat{{\rm N}}_{1}({\bf p})\rangle$ of DQ's with the momenta ${\bf p}$ 
 is given by
\begin{equation}
\tilde{{\rm N}}_{1}({\bf p})=\frac{1}{\exp[\beta(\varepsilon_q({\bf p})-\tilde{\mu})]-1}.
\label{42}
\end{equation}
The average number $\tilde{{\rm N}}_{0}=\langle \hat{{\rm N}}_{0}\rangle$ of DQ's for the discrete energy spectrum  is
\begin{equation}
\tilde{{\rm N}}_{0}=\sum_{n_0}\sum_{n_1}\sum_{n_2} \sum_{n_3}\frac{\delta_{\{n\}}}{\exp[\beta(E_{\{n\}}-\tilde{\mu})]-1}.
\label{43}
\end{equation}
Thus the full average number $\tilde{{\rm N}}_q$ of DQ's is given by
\begin{equation}
\tilde{{\rm N}}_q=\tilde{{\rm N}}_{0}+\tilde{{\rm N}}_{1},~~~ \tilde{{\rm N}}_{1}=\sum_{{\bf p}}\tilde{{\rm N}}_{1}({\bf p}).
\label{44}
\end{equation}
The average number $\tilde{{\rm N}}_{c}$ of DQ's in the ground state (with the energy $E_0$) is 
\begin{equation}
\tilde{{\rm N}}_{c}=\frac{1}{\exp[\beta(E_0-\tilde{\mu})]-1}.
\label{45}
\end{equation}

The density of DQ's in the ground state is $\tilde{n}_c={\rm Lim}(\tilde{{\rm N}}_c/{\rm V})$ where ${\rm Lim}$ denotes the thermodynamical limit (${\rm N},{\rm V}\rightarrow \infty$ for ${\rm N}/{\rm V}=n={\rm constant}$). Here ${\rm N}$ is the full number of atoms in the volume ${\rm V}$.
It follow from Eqs. (\ref{42}) and (\ref{43}) (see also the Appendix C) that the full density of DQ's is
 \begin{eqnarray}
\frac{\tilde{{\rm N}}_q}{{\rm V}}=\left(\frac{M_q}{2\pi\hbar^2\beta}\right)^{3/2}\zeta_{3/2}(\exp(\beta\tilde{\mu}))
\nonumber\\ \noalign{\vskip3pt}      +\frac{1}{{\rm V}}\sum_{\{n\}}\frac{\delta_{\{n\}}}{\exp[\beta (E_{\{n\}}-\tilde{\mu})]-1},
\label{46}
\end{eqnarray}
where the first and the second terms are $\tilde{{\rm N}}_{1}/{\rm V}$ and $\tilde{{\rm N}}_{0}/{\rm V}$ respectively. The function $\zeta_s(z)$ is the polylogarithm which is the generalisation of the Riemann zeta-function given by
\[ \zeta_s(z)\equiv{\rm Li}_{s}(z)=\sum_{n=1}^{\infty}\frac{z^n }{n^{s}}.
\]

 The chemical potential $\tilde{\mu}$ is the volume depended function. We also define the chemical potential $\mu={\rm Lim}~\tilde {\mu}$ which is the limiting case of the chemical potential $\tilde{\mu}$.
 It follows from Eq. (\ref{46}) that $\mu=E_0$ for low temperatures when ${\rm T}<{\rm T}_c$.
 The critical temperature ${\rm T}_c$ for the fixed density $\tilde{n}_q$ follows from the equation, 
 \begin{equation}
\tilde{n}_q=\left(\frac{M_qk_B{\rm T}_c}{2\pi\hbar^2}\right)^{3/2}\zeta_{3/2}(z_c),~~~z_c=\exp\left(\frac{E_0}{k_B{\rm T}_c} \right).
\label{47}
\end{equation}

The  Eq. (\ref{46}) also yields $\tilde{n}_q=\tilde{n}_0+\tilde{n}_1$, where $\tilde{n}_0={\rm Lim}(\tilde{{\rm N}}_{0}/{\rm V})$ and $\tilde{n}_1={\rm Lim}(\tilde{{\rm N}}_{1}/{\rm V})$, and the density $\tilde{n}_1$ for the case ${\rm T}<{\rm T}_c$ is given by equation,
\begin{equation}
\tilde{n}_1=\left(\frac{M_q}{2\pi\hbar^2\beta}\right)^{3/2}\zeta_{3/2}(\exp(\beta E_0)).
\label{48}
\end{equation}

We present below the theorem which defines the volume depended chemical potential $\tilde{\mu}$ in the range of temperatures ${\rm T}\leq{\rm T}_c$. It is important that this equation for the volume depended chemical potential $\tilde{\mu}$ leads to correct thermodynamical limit for the condensed fraction. 

{\it Theorem:  The chemical potential $\tilde{\mu}$ for the arbitrary fixed density $\tilde{n}_q$ in the range of temperatures ${\rm T}<{\rm T}_c$ is 
\begin{equation}
\tilde{\mu}(\tilde{n}_q,{\rm T},{\rm V})=\mu-k_B{\rm T}\left(\tilde{{\rm N}}_c^{-1}-\frac{1}{2}\tilde{{\rm N}}_c^{-2}+\frac{1}{3}\tilde{{\rm N}}_c^{-3} -... \right),
\label{49}
\end{equation}
where $\mu=E_0$ and $\tilde{{\rm N}}_{c}=(\tilde{n}_q-\tilde{n}_1){\rm V}$. It is assumed the volume ${\rm V}$ is enough large that $(\tilde{n}_q-\tilde{n}_1){\rm V}\gg1$. The densities $\tilde{n}_q$ and $\tilde{n}_1$ are given by Eqs. (\ref{47}) and (\ref{48}) where $\tilde{n}_q=\tilde{n}_0+\tilde{n}_1$, and 
$\tilde{n}_0=\tilde{n}_c>0$ for ${\rm T}<{\rm T}_c$.

The chemical potential $\tilde{\mu}$ for the critical temperature ${\rm T}_c$ has the form $\tilde{\mu}=E_0-\varepsilon$, where $\varepsilon>0$. It is assumed that the limit $\varepsilon\rightarrow 0$ always follows after the thermodynamical limit}. 

First we show that the Eq. (\ref{49}) yields the correct density 
$\tilde{n}_0$ in the thermodynamical limit. The Eqs. (\ref{43}) and (\ref{49}) lead to the next equations,
\begin{eqnarray}
\frac{\tilde{{\rm N}}_{0}}{{\rm V}}=\sum_{\{n\}}\frac{\delta_{\{n\}}{\rm V}^{-1}}{\exp[\beta (E_{\{n\}}-\tilde{\mu})]-1} \simeq \frac{{\rm V}^{-1}}{\exp[\beta(E_0-\tilde{\mu})]-1}
\nonumber\\ \noalign{\vskip3pt}    \simeq \frac{{\rm V}^{-1}}{\exp[1/(\tilde{n}_q-\tilde{n}_1){\rm V}]-1}\simeq \tilde{n}_q-\tilde{n}_1,~~~~~~~~~~~~~
\label{50}
\end{eqnarray}
where we use the decomposition $\exp(1/X)= 1+X^{-1}+(1/2)X^{-2}+...$ for $X=(\tilde{n}_q-\tilde{n}_1){\rm V}\gg 1$. Thus for the range of temperatures ${\rm T}<{\rm T}_c$ the Eqs. (\ref{48}) and (\ref{50}) yield in the thermodynamical limit the equation,
\begin{equation}
\tilde{n}_q=\tilde{n}_0+\left(\frac{M_q}{2\pi\hbar^2\beta}\right)^{3/2}\zeta_{3/2}(\exp(\beta E_0)).
\label{51}
\end{equation}

The Eq. (\ref{45}) yields the necessary condition for the volume depended chemical potential $\tilde{\mu}$. It follows from Eq. (\ref{45}) the next decomposition,
 \begin{eqnarray}
\tilde{\mu}=E_0-k_B{\rm T}~{\rm ln}\left(1+\tilde{{\rm N}}_c^{-1})\right)
\nonumber\\ \noalign{\vskip3pt}   =E_0-k_B{\rm T}\left(\tilde{{\rm N}}_c^{-1}-\frac{1}{2}\tilde{{\rm N}}_c^{-2} +\frac{1}{3}\tilde{{\rm N}}_c^{-3} -... \right),
\label{52}
\end{eqnarray}
where $\tilde{{\rm N}}_{c}\gg 1$. Moreover 
the Eqs. (\ref{43}) and (\ref{45}) yield the relation $\tilde{n}_0=\tilde{n}_c$. Thus the Eq. (\ref{49}) is valid when the condition $\tilde{{\rm N}}_{c}=(\tilde{n}_q-\tilde{n}_1){\rm V}\gg1$ is satisfied. 

It follows from Eq. (\ref{51}) that the full density of DQ's is $\tilde{n}_q=\tilde{n}_0+\tilde{n}_1$ where $\tilde{n}_0=\tilde{n}_c>0$ for the range of temperatures ${\rm T}<{\rm T}_c$, and $\tilde{n}_0=\tilde{n}_c=0$ when ${\rm T}\geq{\rm T}_c$. 
Hence the DQ's condense in the ground state with the energy $E_0$ for the range of temperatures ${\rm T}<{\rm T}_c$. 

The Eq. (\ref{42}) yields the excitation density $\tilde{n}_{1}$ as
\begin{equation}
\tilde{n}_{1}=\frac{1}{(2\pi\hbar)^3}\int\frac{1}{\exp[\beta(\varepsilon_q({\bf p})-E_0)]-1}d{\bf p}.
\label{53}
\end{equation}
We note that for the temperature region ${\rm T}_p<{\rm T}< {\rm T}_{\lambda}$ with ${\rm T}_p\simeq 1$ the inequality $e^{-\beta\mu}=e^{\beta|E_0|}\gg 1$ is satisfied because the ground state energy of the DQ's is $E_0/k_B=-8.65~ {\rm K}$ (see Section II). 
The Eq. (\ref{53}) for the condition $e^{\beta|E_0|}\gg 1$ has the form  $\tilde{n}_{1}=\int f_q({\bf p})d{\bf p}$ where the momentum distribution of diatomic quasiparticles is 
\begin{equation}
f_q({\bf p})=\frac{1}{(2\pi\hbar)^3}\exp[-\beta(\varepsilon_q({\bf p})-E_0)].
\label{54}
\end{equation}
The integration in Eq. (\ref{53}) for the conditions $|\epsilon_q|\ll |E_0|$ and $e^{\beta|E_0|}\gg 1$ leads to the excitation density of the DQ's given by
\begin{equation}
\tilde{n}_{1}({\rm T},\rho)=\frac{1}{\lambda_q^3({\rm T})}\exp\left(-\frac{|E_0(\rho)|}{k_B{\rm T}}\right),
\label{55}
\end{equation}
where the thermal wavelength $\lambda_q({\rm T})$ is
\begin{equation}
\lambda_q({\rm T})=\left(\frac{2\pi\hbar^2}{M_qk_B{\rm T}}\right)^{1/2}.
\label{56}
\end{equation}
The Eq. (\ref{55}) also follows from Eq. (\ref{48}) when the condition $e^{\beta|E_0|}\gg 1$ is satisfied.
 The relation, $\tilde{n}_0=\tilde{n}_c$, and the Eq. (\ref{51}) yield the condensate density as
\begin{equation}
\tilde{n}_0({\rm T},\rho)=\tilde{n}_c({\rm T},\rho)=\tilde{n}_q(\rho)-\tilde{n}_{1}({\rm T},\rho).
\label{57}
\end{equation}
Hence the equation $\tilde{n}_{1}=\tilde{n}_q$ is satisfied at the critical temperature ${\rm T}_c$.

The Eq. (\ref{55}) (for the temperature ${\rm T}_c$) and Eq. (\ref{33}) yield the full density of diatomic quasiparticles as
\begin{equation}
\tilde{n}_q(\rho)=e^{-4}\left(\frac{M_qk_B{\rm T}_c(\rho)}{2\pi\hbar^2}\right)^{3/2}=e^{-4}\left(\frac{M_q|E_0(\rho)|}{8\pi\hbar^2}\right)^{3/2},
\label{58}
\end{equation}
where the functions ${\rm T}_c(\rho)$ and $|E_0(\rho)|$ are presented by Eqs. (\ref{35}) and (\ref{34}) respectively. We emphasise that the Eq. (\ref{58}) yields $\tilde{n}_q/n\ll 1$, where $n={\rm Lim}~ {\rm N}/{\rm V}$ is the density of atoms in liquid ${\rm ^4He}$. Thus the necessary condition for the theory based on DQ's concept is satisfied.

The Eq. (\ref{58}) leads to the critical temperature for BEC in liquid ${\rm ^4He}$ as
\begin{equation}
{\rm T}_c(\rho)=\frac{2\pi e^{8/3}\hbar^2\tilde{n}_q(\rho)^{2/3}}{k_BM_q}=\frac{|E_0(\rho)|}{4k_B}.
\label{59}
\end{equation}

 We note that the energy density, ${\cal U}_q={\rm Lim}({\rm U}_q/{\rm V})$, (see Appendix C) of diatomic quasiparticles can be written by Eq. (\ref{c8}) in the form,
\begin{equation}
{\cal U}_q({\rm T},\rho)=\frac{3}{2}k_B{\rm T}\tilde{n}_{1}({\rm T},\rho)+E_0(\rho)(\tilde{n}_q(\rho)-\tilde{n}_{1}({\rm T},\rho)).
\label{60}
\end{equation}
The entropy density of the diatomic quasiparticles follows by Eq. (\ref{c9}) as
\begin{equation}
{\cal S}_q({\rm T},\rho)={\rm Lim}\frac{S_q}{{\rm V}}=\left(\frac{5}{2}k_B+\frac{|E_0(\rho)|}{{\rm T}}\right)\tilde{n}_{1}({\rm T},\rho).
\label{61}
\end{equation}
The Eqs. (\ref{60}) and (\ref{61}) lead to free energy density, ${\cal F}_q={\cal U}_q-{\rm T}{\cal S}_q$, of diatomic quasiparticles as
\begin{equation}
{\cal F}_q({\rm T},\rho)=E_0(\rho)\tilde{n}_q(\rho)-k_B{\rm T}\tilde{n}_{1}({\rm T},\rho).
\label{62}
\end{equation}

In the paper \cite{Krug} are found the equations, $\Delta=-E_0$ and $\mu_{\rm rot}= - \Delta$, where $\Delta$ and 
$\mu_{\rm rot}$ are the roton gap and the roton chemical potential respectively. Furthermore the theorem proven in this section yields the equation ${\rm Lim}~\tilde{\mu}=\mu=E_0$ where the ground state energy of DQ's is given by 
Eq. (\ref{19}). These equations lead to the general relation $\mu=\mu_{\rm rot}=E_0$ for the temperature region 
${\rm T}_p \leq {\rm T} \leq {\rm T}_\lambda$. This thermodynamic relation for the chemical potentials $\mu$ and $\mu_{\rm rot}$
is always satisfied when the DQ's and rotons exist in liquid ${\rm ^4He}$.

\section{Superfluid and BEC fractions in liquid ${\rm ^4He}$ for temperatures ${\rm T}_p \leq {\rm T} \leq {\rm T}_\lambda$}

In this section we derive the BEC and superfluid fractions in liquid ${\rm ^4He}$ for the temperatures ${\rm T}_p \leq {\rm T} \leq {\rm T}_\lambda$ with ${\rm T}_p\simeq 1~{\rm K}$. We also show that these superfluid and BEC fractions are in good agreement with experimental data and Monte Carlo simulations \cite{Cep,Mas}.
We note that in the temperature interval ${\rm T}_p \leq {\rm T} \leq {\rm T}_\lambda$ exist three type of quasiparticles:  diatomic quasiparticles, rotons, and phonons. However we show that the superfluid and BEC fractions in liquid ${\rm ^4He}$ are completely defined by the thermodynamical functions for the diatomic quasiparticles in the temperature region ${\rm T}_p \leq {\rm T} \leq {\rm T}_\lambda$. 

The full mass density of DQ's and the mass densities for the condensed and excited quasiparticles are given as $\rho_q=M_q\tilde{n}_q$,  $\rho_{0}=M_q\tilde{n}_{0}$, and $\rho_{1}=M_q\tilde{n}_{1}$ respectively.
The condensate fraction in liquid Bose fluid for the temperature region ${\rm T}_p \leq {\rm T} \leq {\rm T}_\lambda$ is given as $\rho_0/\rho=(\tilde{n}_q-\tilde{n}_{1})M_q/\rho$. The Eqs. (\ref{55}), (\ref{58}) and (\ref{59}) and the relation ${\rm T}_c={\rm T}_\lambda$ yield
\begin{equation}
\frac{\rho_0}{\rho}=f({\rho})[1-\Phi({\rm T},\rho)],
\label{63}
\end{equation}
\begin{equation}
\frac{\rho_{1}}{\rho}=f({\rho})\Phi({\rm T},\rho).
\label{64}
\end{equation}
The functions $f({\rho})$ and $\Phi({\rm T},\rho)$  are given by
\begin{equation}
f({\rho})=\frac{M_q}{e^4\rho}\left(\frac{M_qk_B{\rm T}_\lambda(\rho)}{2\pi\hbar^2}\right)^{3/2},
\label{65}
\end{equation}
\begin{equation}
\Phi({\rm T},\rho)=\left(\frac{{\rm T}}{{\rm T}_\lambda(\rho)}\right)^{3/2}\exp\left[4\left(1-\frac{{\rm T}_\lambda(\rho)}{{\rm T}}\right)\right],
\label{66}
\end{equation}
where the lambda transition temperature ${\rm T}_\lambda(\rho)$ is the function of density  given in Eq. (\ref{35}). 
We emphasize that the excitation mass density $\rho_{1}=\rho_q-\rho_0$ is connected to full density $\rho_q$ of diatomic quasiparticles. One can also introduce the full excitation mass density given by $\rho_{\rm ex}=\rho-\rho_0$ then the full excitation fraction is
\begin{equation}
\frac{\rho_{\rm ex}}{\rho}=1-f({\rho})[1-\Phi({\rm T},\rho)].
\label{67}
\end{equation}
The Eqs. (\ref{63}),  (\ref{67}) lead to the relation $\rho_{\rm ex}/\rho+\rho_{0}/\rho=1$.

The energy density of diatomic quasiparticles can be written in the normalised form $\tilde{{\cal U}}_q={\cal U}_q({\rm T})/{\cal U}_q({\rm T}_\lambda)$ where the energy density ${\cal U}_q({\rm T})$ is given by Eq. (\ref{60}) and we have ${\cal U}_q({\rm T}_\lambda)=(3/2)k_B{\rm T}_\lambda \tilde{n}_q$. This leads to the normalised energy density as

\begin{equation}
\tilde{{\cal U}}_q({\rm T},\rho)=\left(\frac{8}{3} +\frac{{\rm T}}{{\rm T}_\lambda(\rho)}\right)\Phi({\rm T},\rho)-\frac{8}{3}.
\label{68}
\end{equation}
The normalised entropy density of diatomic quasiparticles is $\tilde{{\cal S}}_q={\cal S}_q({\rm T})/{\cal S}_q({\rm T}_\lambda)$ where the entropy density ${\cal S}_q({\rm T})$ is given by Eq. (\ref{61}) and we have ${\cal S}_q({\rm T}_\lambda)=(13/2)k_B\tilde{n}_q$. Hence the normalised entropy density of diatomic quasiparticles is
\begin{equation}
\tilde{{\cal S}}_q({\rm T},\rho)=\left[\frac{5}{13} +\frac{8}{13}\left( \frac{{\rm T}_\lambda(\rho)}{{\rm T}} \right)\right]\Phi({\rm T},\rho).
\label{69}
\end{equation}

We have two different equations for the temperature interval ${\rm T}_p \leq {\rm T} \leq {\rm T}_\lambda$ given by
\begin{equation}
\rho_q=\rho_{0}+\rho_{1},~~~\rho=\rho_{s}+\rho_{n},
\label{70}
\end{equation}
where second equation is written for the superfluid $\rho_{s}$ and normal $\rho_{n}$ mass densities.
We can also write for this temperature region the next equation $\rho_{1}=\alpha(\rho) \rho_n$ where it is assumed that $\alpha(\rho)$ is some function of density. At the temperature ${\rm T} = {\rm T}_\lambda$ we have the relations $\rho_{1}=\rho_q$ and $\rho_n=\rho$ which yield $\rho_q=\alpha(\rho) \rho$.
The Eq. (\ref{70}) and the relations $\rho_{1}=\alpha(\rho) \rho_n$ and $\rho_q=\alpha(\rho) \rho$ lead to the equation $\rho_{0}=\alpha(\rho) \rho_{s}$. Thus for the temperature region ${\rm T}_p \leq {\rm T} \leq {\rm T}_\lambda$ we have the next equations, 
\begin{equation}
\frac{\rho_{1}}{\rho}=\alpha(\rho)\frac{\rho_{n}}{\rho},~~~\frac{\rho_{0}}{\rho}=\alpha(\rho)\frac{\rho_{s}}{\rho},
\label{71}
\end{equation}
where $\alpha(\rho)=\rho_q(\rho)/\rho$. The relation $\rho_{\rm ex}/\rho=1-\rho_{0}/\rho$ and Eq. (\ref{71}) also lead to the equation,
\begin{equation}
\frac{\rho_{\rm ex}}{\rho}=1-\alpha(\rho)+\alpha(\rho)\frac{\rho_{n}}{\rho}.
\label{72}
\end{equation}

The Eq. (\ref{64}) for the temperature ${\rm T}={\rm T}_\lambda$ yields $\rho_q(\rho)/\rho=f(\rho)$ and hence we have the next relation $\alpha(\rho)\equiv f(\rho)$.
It follows that the Eqs. (\ref{35}), (\ref{65}) and relation $\alpha(\rho)\equiv f(\rho)$ yield the function $\alpha(\rho)$ as
\begin{equation}
\alpha(\rho)=\left(\frac{M_q\epsilon}{2\pi\hbar^2}\right)^{3/2} \frac{M_q} {e^4\rho}\left[\frac{1}{4}-\left(\frac{\rho}{\rho_a}\right)^{2}+\left(\frac{\rho}{\rho_a}\right)^{4}\right]^{3/2},
\label{73}
\end{equation}
where $\rho_a=6ma^3/\pi r_0^3$. 
The Eqs. (\ref{63}), (\ref{64}), (\ref{71}) and relation $\alpha(\rho)\equiv f(\rho)$ lead to the normal and superfluid fractions as 
\begin{equation}
\frac{\rho_n}{\rho}=\Phi({\rm T},\rho),~~~\frac{\rho_s}{\rho}=[1-\Phi({\rm T},\rho)],
\label{74}
\end{equation}
where the function $\Phi({\rm T},\rho)$ is given in Eq. (\ref{66}).
\begin{figure}
\includegraphics[width=9cm,trim=0mm 16mm 0mm 0mm]{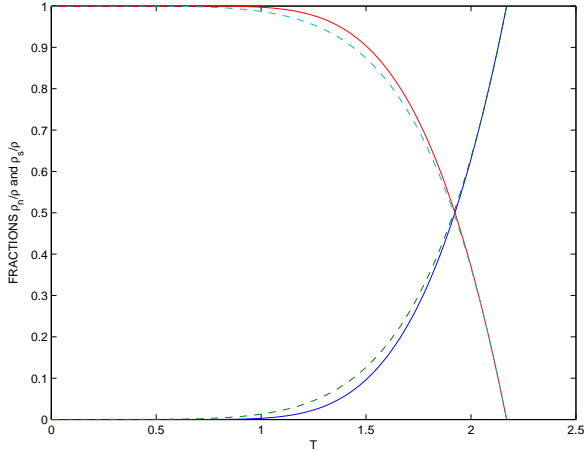}
\caption{(Color online) Normal and superfluid fractions given by Eq. (\ref{74}) (theory-solid line) and fit by Eq. (\ref{75}) (experimental data-dashed line) for temperature region ${\rm T}_p \leq {\rm T} \leq {\rm T}_\lambda$ and density $\rho=0.145$ ${\rm g~cm^{-3}}$.} 
\label{fig:1}
\end{figure}
\begin{figure}
\includegraphics[width=9cm,trim=0mm 16mm 0mm 0mm]{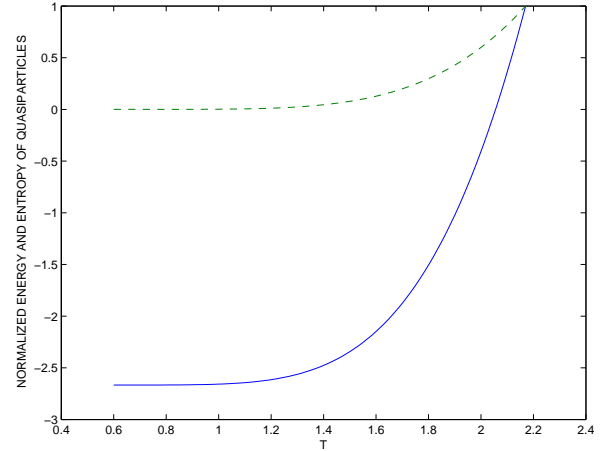}
\caption{(Color online) Normalised energy (solid line) and entropy (dashed line) of diatomic quasiparticles given by Eqs. (\ref{68}) and (\ref{69}) for density $\rho=0.145$ ${\rm g~cm^{-3}}$.} 
\label{fig:2}
\end{figure}
\begin{figure}
\includegraphics[width=9cm,trim=0mm 16mm 0mm 0mm]{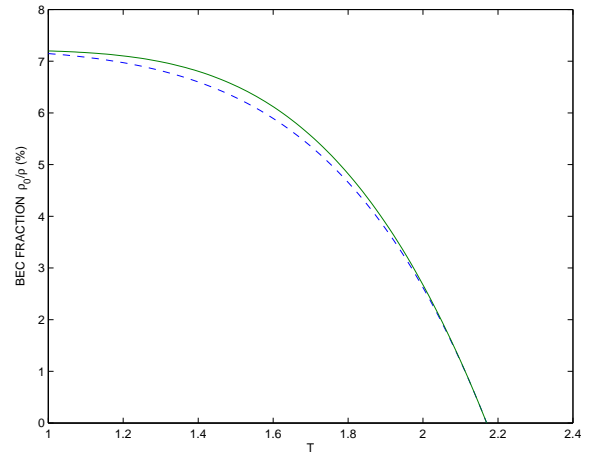}
\caption{(Color online) Condensate fraction given by Eq. (\ref{63}) (theory-solid line) and fit by Glide \cite {Gl} (experimental data-dashed line) for temperature region ${\rm T}_p \leq {\rm T} \leq {\rm T}_\lambda$ ($\rho=0.145$ ${\rm g~cm^{-3}}$).}
 \label{fig:3}
\end{figure}
\begin{figure}
\includegraphics[width=9cm,trim=0mm 16mm 0mm 0mm]{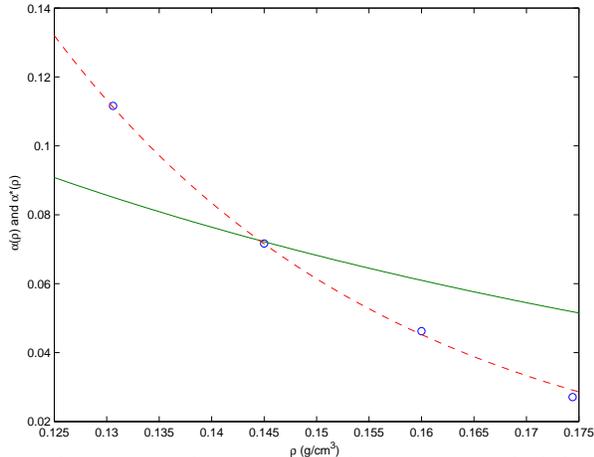}
\caption{(Color online) Function $\alpha(\rho)$ given by Eq. (\ref{73}) (solid line) for the region ${\rm T}_p \leq {\rm T} \leq {\rm T}_\lambda$, and function $\alpha^{*}(\rho)$ given by Eq. (\ref{85}) (dashed line)) for temperature region ${\rm T} \leq 0.5~{\rm K}$. Condensate fraction in liquid ${\rm ^4He}$ for zero temperature Ref. \cite{Moro} (circles).} 
\label{fig:4}
\end{figure}
The experimental data for the normal and superfluid fractions in liquid $^4$He are well approximated for the range of temperatures ${\rm T}_p \leq {\rm T} \leq {\rm T}_\lambda$ at saturated vapour pressure (SVP) by equations,
\begin{equation}
\frac{\rho_n}{\rho}=\left(\frac{{\rm T}}{{\rm T}_\lambda}\right)^{5.6},~~~\frac{\rho_s}{\rho}=1-\left(\frac{{\rm T}}{{\rm T}_\lambda}\right)^{5.6}.
\label{75}
\end{equation}

The normal and superfluid fractions given by Eqs. (\ref{74}) (the theory) and (\ref{75}) (the experimental data) are shown in Fig. 1. In Fig. 2 we present the normalised energy and entropy of diatomic quasiparticles given by Eqs. (\ref{68}) and (\ref{69}) respectively. 
In Fig. 3  we present the condensate fraction given by Eq. (\ref{63}) (the theory) and the fit of the observed data for the condensate fraction by Glyde \cite {Gl} which at SVP is  $\rho_0/\rho=0.0725[1-({\rm T}/{\rm T}_\lambda)^{5.5}]$.

Thus the Fig. 1 and Fig. 3 demonstrate good agreement of the theoretical equations with the experimental data for superfluid and BEC fractions for the range of temperatures ${\rm T}_p \leq {\rm T} \leq {\rm T}_\lambda$ at SVP.

\section{Superfluid and BEC fractions in liquid ${\rm ^4He}$ for low temperatures}

In the temperature interval ${\rm T} \leq 0.5~{\rm K}$ the most important excitations are the phonons. 
The phonons energy is a linear function of the wave number $k$: $E_k=u_1\hbar k$.
This relation yields the equation for free energy per unit volume as
\begin{equation}
{\cal F}({\rm T},\rho)={\cal F}_0(\rho)-\frac{\pi^2 k_B^4{\rm T}^4}{90\hbar^3u_1^3},
\label{76}
\end{equation}
where $u_1$ is the first sound, ${\cal F}_0(\rho)={\cal E}_0(\rho)\rho/m$ and ${\cal E}_0(\rho)$ is the ground state energy per one particle at zero temperature. The second sound in two-fluid hydrodynamics is given by
\begin{equation}
u_2=\sqrt{\frac{\rho_s s^2}{\rho_n(\partial s/\partial {\rm T})_\rho}},
\label{77}
\end{equation}
where $s={\cal S}/\rho$, and ${\cal S}=-\partial{\cal F}/\partial{\rm T}$ is the entropy per unit volume.
The entropy $s$ follows from Eq. (\ref{76}) as
\begin{equation}
s=\Lambda {\rm T}^3,~~~\Lambda=\frac{2\pi^2 k_B^4}{45\hbar^3u_1^3\rho},
\label{78}
\end{equation}
where the first sound $u_1$ does not depend on the temperature for the region ${\rm T}<0.5~{\rm K}$.

We seek the normal and superfluid fractions for liquid ${\rm ^4He}$ at low temperatures ${\rm T}\leq 0.5~{\rm K}$ in the form:
\begin{equation}
\frac{\rho_n}{\rho}=\gamma(\rho){\rm T}^\nu,~~~\frac{\rho_s}{\rho}=1-\gamma(\rho){\rm T}^\nu.
\label{79}
\end{equation}
The Eqs. (\ref{76})-(\ref{79}) yield the second sound as
\begin{equation}
u_2=\sqrt{\frac{\Lambda}{3\gamma}({\rm T}^{4-\nu}-\gamma{\rm T}^4)}.
\label{80}
\end{equation}
There are only three different cases for the parameter $\nu$:

(1) $\nu<4$, then $u_2\rightarrow 0$ for ${\rm T}\rightarrow 0$,

(2) $\nu>4$, then $u_2\rightarrow \infty$ for ${\rm T}\rightarrow 0$,

(3) $\nu=4$, then  $u_2\rightarrow \sqrt{\Lambda/3\gamma}$ for ${\rm T}\rightarrow 0$.
 
 We choose the case (3) with $\nu=4$ because $u_2\neq 0$ and $u_2\neq \infty$ at zero temperature. Moreover the limiting value for second sound at ${\rm T}\rightarrow 0$ is $u_2=u_1/\sqrt{3}$ because the energy of the phonon excitations is a linear function of the wave number.
It follows from Eq. (\ref{80}) that the second sound for ${\rm T}\leq 0.5~{\rm K}$ is  
\begin{equation}
u_2=\frac{u_1}{\sqrt{3}}\left(1-\gamma(\rho){\rm T}^4\right)^{1/2},~~~\gamma(\rho)=\frac{\Lambda}{u_1^2}.
\label{81}
\end{equation}
The function $\gamma(\rho)$ is given by Eqs. (\ref{78}) and (\ref{81}) as
\begin{equation}
\gamma(\rho)=\frac{2\pi^2 k_B^4}{45\hbar^3u_1^5\rho}.
\label{82}
\end{equation}
The Eqs. (\ref{79}) and (\ref{82}) lead to the normal and superfluid fractions of liquid ${\rm ^4He}$ for ${\rm T}\leq 0.5~{\rm K}$ (see also \cite{Lan}) as
\begin{equation}
\frac{\rho_n}{\rho}=\left(\frac{2\pi^2 k_B^4}{45\hbar^3u_1^5\rho}\right){\rm T}^4,~~~\frac{\rho_s}{\rho}=1-\left(\frac{2\pi^2 k_B^4}{45\hbar^3u_1^5\rho}\right){\rm T}^4.
\label{83}
\end{equation}
We assume the relation $\rho_0=\alpha^{*}(\rho)\rho_s$ for the temperature interval ${\rm T}\leq 0.5~{\rm K}$. In this case the equations $\rho=\rho_0+\rho_{\rm ex}$ and $\rho=\rho_s+\rho_n$ yield
\begin{equation}
\frac{\rho_{0}}{\rho}=\alpha^{*}(\rho)\frac{\rho_{s}}{\rho},~~~\frac{\rho_{\rm ex}}{\rho}=1-\alpha^{*}(\rho)+\alpha^{*}(\rho)\frac{\rho_{n}}{\rho}.
\label{84}
\end{equation}

These equations are similar to Eqs. (\ref{71}) and (\ref{72}). However the functions $\alpha(\rho)$ and $\alpha^{*}(\rho)$ are given for different temperature regions. It follows from the Eqs. (\ref{83}) and (\ref{84}) that for zero temperature the condensate fraction is given by $\rho_0/\rho=\alpha^{*}(\rho)$. 

The function $\alpha^{*}(\rho)$ can be found by Monte Carlo simulations of the condensate fraction in liquid ${\rm ^4He}$ at zero temperature. We choose the function $\alpha^{*}(\rho)$ in the form,
\begin{equation}
\alpha^{*}(\rho)=C\exp(-\kappa_0\rho),
\label{85}
\end{equation}
which is relevant to Feynman approximation \cite{Fe,Pen} of the ground-state wave function. However such treatment leads to a rough estimate for the constants in Eq. (\ref{85}). 
The fitting the Eq. (\ref{85}) by data from Moroni et al. \cite{Moro} yields the constants as $C=6.05$ and $\kappa_0=30.6$ ${\rm cm^{3}~g^{-1}}$.

 The functions $\alpha(\rho)$ and $\alpha^{*}(\rho)$ given by Eqs. (\ref{73}) and (\ref{85}) are shown in Fig. 4. This figure demonstrates that the functions $\alpha(\rho)$ and $\alpha^{*}(\rho)$ are different. However at low pressure when the density is $\rho\simeq 0.145$ ${\rm g~cm^{-3}}$ we have the relation $\alpha(\rho)\simeq \alpha^{*}(\rho)$.
The Eqs. (\ref{83}) and (\ref{84}) yield the condensate and excitation fractions as 
\begin{equation}
\frac{\rho_0}{\rho}=\alpha^{*}(\rho)-\left(\frac{2\pi^2 k_B^4\alpha^{*}(\rho)}{45\hbar^3u_1^5\rho}\right){\rm T}^4,
\label{86}
\end{equation}
\begin{equation}
\frac{\rho_{\rm ex}}{\rho}=1-\alpha^{*}(\rho)+\left(\frac{2\pi^2 k_B^4\alpha^{*}(\rho)}{45\hbar^3u_1^5\rho}\right){\rm T}^4.
\label{87}
\end{equation}

We note that for the range of temperatures ${\rm T}\leq 0.5~{\rm K}$ the phonon velocity $u_1$ is a linear function of density $u_1=\sigma_0+\sigma_1\rho$  \cite{Kr}. 
The Eq. (\ref{86}) for zero temperature leads to the condensate fraction of liquid ${\rm ^4He}$ as 
$\rho_0/\rho=\alpha^{*}(\rho)$ which at SVP is $\rho_0/\rho\simeq \alpha(\rho)=7.22~\%$. 
Recent measurements show that for low temperatures at SVP the condensate fraction is $7.25\pm 0.75~\%$ Ref. \cite{Gl}. The diffusion Monte Carlo simulations \cite{Moro} for zero temperature gives the condensate fraction as $7.17~\%$. These results also agree for both Glyde et al. \cite{D2} and Snow et al. \cite{Sno}. 

The theory also leads to the bound temperature ${\rm T}_p$ given as a function of density. The Eqs. (\ref{74}) and (\ref{83}) with ${\rm T}={\rm T}_p$ yield the equation for the bound temperature ${\rm T}_p$ as
\begin{equation}
\Phi({\rm T}_p,\rho)=\gamma(\rho){\rm T}_p^4.
\label{88}
\end{equation}
More simple equation for the temperature ${\rm T}_p$ follows by 
Eqs. (\ref{75}) and (\ref{83}). One can also use in Eq. (\ref{75}) the power $\nu=11/2$ which does not violate the accuracy for the bound temperature ${\rm T}_p$. In this case we have the next equation,
\begin{equation}
\left(\frac{{\rm T}_p}{{\rm T}_\lambda}\right)^{11/2}=\left(\frac{2\pi^2 k_B^4}{45\hbar^3u_1^5\rho}\right){\rm T}_p^4,
\label{89}
\end{equation}
which yields 
\begin{equation}
{\rm T}_p=\left(\frac{2\pi^2 k_B^4{\rm T}_\lambda^{11/2}}{45\hbar^3u_1^5\rho}\right)^{2/3}.
\label{90}
\end{equation}
Hence the bound temperature at SVP is ${\rm T}_p\simeq 1~{\rm K}$. This value also agrees with the experimental data presented in Ref.\cite{Kler} and \cite{EnH}.

\section{Conclussions}

 In the present theory the lambda transition and the Bose-Einstein condensation in liquid ${\rm ^4He}$ are described by diatomic quasiparticles. The theory demonstrates that in liquid ${\rm ^4He}$ for the temperature region $1~{\rm K}\leq{\rm T}\leq {\rm T}_\lambda$ the diatomic quasiparticles macroscopically populate the ground state which leads to BEC in liquid ${\rm ^4He}$. This approach yields the lambda transition temperature ${\rm T}_\lambda=2.16~{\rm K}$ which is very close to experimental lambda temperature ${\rm T}_\lambda=2.17~{\rm K}$. The Fig. 1 and Fig. 3 also demonstrate good agreement of the theoretical superfluid and BEC fractions with experimental observations. These results confirm the existence of DQ's in liquid ${\rm ^4He}$ at low temperatures. It is also shown the connection between BEC and superfluidity phenomena in the temperature intervals $1~{\rm K}\leq{\rm T}\leq {\rm T}_\lambda$ and ${\rm T} \leq 0.5~{\rm K}$ by scaling laws given in Eqs. (\ref{71}), (\ref{72}) and (\ref{84}). 
 
 The theory demonstrates that the BEC in noninteracting Bose gas and Bose fluid have different nature. Indeed the BEC in Bose fluid is connected with the coupling of strong interacting atoms which form the diatomic quasiparticles in liquid ${\rm ^4He}$ for the temperature region ${\rm T}_p\leq{\rm T}\leq {\rm T}_\lambda$. 

This approach can also be extended to the hydrodynamics of Bose fluid.
We note that the two-fluid hydrodynamics is self-consistent as phenomenological theory of superfluidity where nevertheless does not explicitly introduced the idea of BEC \cite{Leg}. However it is shown in this paper that the BEC and superfluidity are connected with the existence of the diatomic quasiparticles in liquid ${\rm ^4He}$. The present theory describes to good accuracy the lambda transition temperature, the Bose-Einstein condensation and superfluidity of liquid ${\rm ^4He}$ at the temperatures below lambda transition and low pressures.

\section*{ACKNOWLEDGMENTS}

The author is grateful to Gerard Milburn and Karen Kheruntsyan for useful discussion of the results of this work. 
 
\appendix

\section{Effective mass of diatomic quasiparticles}

We presented in this Appendix the renormalisation procedure for the mass of diatomic quasiparticles in liquid ${\rm ^4He}$ for the temperature region ${\rm T}_p \leq {\rm T} \leq {\rm T}_\lambda$.

The energy of the excited diatomic quasiparticles per unit volume follows from Eq. (\ref{60}) as 
\begin{equation}
{\cal U}_{1}=\frac{3}{2}k_B{\rm T}\tilde{n}_{1}=\frac{3k_B{\rm T}}{2\lambda_q^3({\rm T})}\exp\left(-\frac{|E_0|}{k_B{\rm T}}\right).
\label{a1}
\end{equation}
The kinetic energy density of the diatomic quasiparticles with the mass $M=2m$ is given by
\begin{equation}
{\cal K}={\rm Lim}\frac{1}{{\rm V}}\sum_{{\bf p}}\frac{\varepsilon({\bf p})}{\exp[\beta(\varepsilon({\bf p})-\mu)]-1},
\label{a2}
\end{equation}
where $\varepsilon({\bf p})={\bf p}^2/2M$. The integration in this equation yields 
\begin{eqnarray}
{\cal K}=\frac{1}{(2\pi\hbar)^3}\int\varepsilon({\bf p})\exp[-\beta(\varepsilon({\bf p})-E_0)]d{\bf p}~~~~~
\nonumber\\ \noalign{\vskip3pt} =\frac{3k_B{\rm T}}{2\lambda^3({\rm T})}\exp\left(-\frac{|E_0|}{k_B{\rm T}}\right),~~~~~~~~~~~
\label{a3}
\end{eqnarray}
where $\lambda({\rm T})$ is the thermal wavelength,
\begin{equation}
\lambda({\rm T})=\left(\frac{2\pi\hbar^2}{Mk_B{\rm T}}\right)^{1/2}.
\label{a4}
\end{equation}
The energy density in Eq. (\ref{a1}) can be written as ${\cal U}_{1}={\cal K}+{\cal V}$ where the potential energy density of the excited quasiparticles in harmonic approximation is given by equation ${\cal V}= {\cal K} $. Thus we have for harmonic approximation the relation,
\begin{equation}
{\cal U}_{1}=2{\cal K}.
\label{a5}
\end{equation}
The substitution of the equations (\ref{a1}) and (\ref{a3}) into equation (\ref{a5}) yields 
\begin{equation}
\frac{1}{\lambda_q^3({\rm T})}= \frac{2}{\lambda^3({\rm T})},
\label{a6}
\end{equation}
where the thermal wavelength $\lambda_q({\rm T})$ is given in Eq. (\ref{56}).
The Eq. (\ref{a6}) leads to the effective (renormalised) mass,
\begin{equation}
M_q=\sigma M,~~~\sigma= 2^{2/3}\simeq 1.587.
\label{a7}
\end{equation}
Hence the renormalised mass of the diatomic quasiparticles is given by $M_q= 2^{5/3}m$.

\section{Vibrational modes in liquid ${\rm ^4He}$}

We consider in this  Appendix the energy of vibrational modes of diatomic quasiparticles. The Hamiltonian for harmonic approximation is given by
\begin{equation}
\hat{{\rm H}}_s=\frac{\hat{p}_s^2}{2M_q}+\frac{1}{2}M_q\omega_s^2\hat{x}_s^2,
\label{b1}
\end{equation}
where $s=1,2,3$.
This Hamiltonian yields the equation,
\begin{equation}
\frac{1}{2}M_q\omega_s^2\langle\hat{x}_s^2\rangle= \langle\frac{\hat{p}_s^2}{2M_q}\rangle,
\label{b2}
\end{equation}
where the average kinetic energy is 
\begin{equation}
\langle\frac{\hat{p}_s^2}{2M_q}\rangle=\frac{1}{2}k_B{\rm T}.
\label{b3}
\end{equation}
The Eqs. (\ref{b2}) and (\ref{b3}) lead to the relation,
\begin{equation}
\frac{\hbar\omega_s}{k_B{\rm T}}= \frac{\hbar}{\sqrt{M_qk_B{\rm T}\langle\hat{x}_s^2\rangle}},
\label{b4}
\end{equation}
where the effective mass is given by $M_q= 2^{5/3}m$ (see Appendix A).
We assume that for the mass density $\rho=0.145$ ${\rm g~cm^{-3}}$ the next relation $\sqrt{\langle\hat{x}_s^2\rangle}\simeq 2d$ is approximately satisfied, where $d\simeq 4.4\cdot 10^{-8}~{\rm cm}$ is the average distance between helium atoms. In this case the Eq. (\ref{b4}) for the critical temperature ${\rm T}_c$ yields the equation $\hbar\omega_s/k_B{\rm T}_c\simeq 1/7$. Thus the inequality $\hbar\omega_s\ll k_B{\rm T}_c$ is satisfied. This inequality also yields $\hbar\omega_k\ll |E_0|$ because the next condition $|E_0|>k_B{\rm T}_c$ is satisfied. The vibrational mode with $s=0$ can be treated similarly.

\section{Thermodynamics of diatomic quasiparticles in helium fluid}

The partition function in Eq. (\ref{41}) for the grand canonical density operator is given by
\begin{eqnarray}
{\rm ln}~\Xi_q=-\sum_{{\bf p}}{\rm ln}(1-z_1e^{-\beta\varepsilon_q({\bf p})})
\nonumber\\ \noalign{\vskip3pt}     -\sum_{n_0}\sum_{n_1}\sum_{n_2} \sum_{n_3}\delta_{\{n\}}{\rm ln}(1-z_0e^{-\beta E_{\{n\}}}),
\label{c1}
\end{eqnarray}
where $z_k=\exp(\beta\mu_k)$ ($k=0,1$). 

The change  
${\rm V}^{-1}\sum_{{\bf p}}\rightarrow (2\pi\hbar)^{-3}\int d{\bf p}$ (for the case ${\rm V}\rightarrow \infty$) in the Eq. (\ref{c1}) yields
\begin{equation}
\frac{{\rm ln}~\Xi_q}{{\rm V}}=\frac{1}{\lambda_q^3}\zeta_{5/2}(z_1)-\frac{1}{{\rm V}}\sum_{{\{n\}}}\delta_{\{n\}}{\rm ln}(1-z_0e^{-\beta E_{{\{n\}}}}),
\label{c2}
\end{equation}
where $\zeta_s(z)\equiv{\rm Li}_{s}(z)$ is the polylogarithm (or generalised zeta-function) and $\lambda_q$ is the thermal wavelength,
\begin{equation}
\zeta_s(z)=\sum_{n=1}^{\infty}\frac{z^n}{n^s},~~~\lambda_q=\left(\frac{2\pi\hbar^2\beta}{M_q}\right)^{1/2}.
\label{c3}
\end{equation}

The density of DQ's for continuos and discrete energy spectrums are
 \begin{equation}
\frac{\tilde{{\rm N}}_1}{{\rm V}}=\frac{1}{\beta}\left(\frac{\partial}{\partial\mu_1}({\rm V}^{-1}{\rm ln}~\Xi_q)\right)_{\mu_k=\tilde{\mu}}=\frac{1}{\lambda_q^3}\zeta_{3/2}(z),
\label{c4}
\end{equation}
\begin{eqnarray}
\frac{\tilde{{\rm N}}_0}{{\rm V}}=\frac{1}{\beta}\left(\frac{\partial}{\partial\mu_0}({\rm V}^{-1}{\rm ln}~\Xi_q)\right)_{\mu_k=\tilde{\mu}}
\nonumber\\ \noalign{\vskip3pt}      =\frac{1}{{\rm V}}\sum_{\{n\}}\frac{\delta_{\{n\}}}{z^{-1}\exp(\beta E_{\{n\}})-1},
\label{c5}
\end{eqnarray}
where $\mu_k=\tilde{\mu}$ ($k=1,2$) and $z=\exp(\beta\tilde{\mu})$. The chemical potential $\tilde{\mu}$ is given in Eq. (\ref{49}).

One can find that the condition $e^{-\beta |E_0|}\leq 0.0186$ is satisfied for ${\rm T}\leq 2.17~{\rm K}$ because $E_0/k_B=-8.65~{\rm K}$. Hence the condition $e^{-\beta |E_0|}\ll 1$ is satisfied when ${\rm T}\leq {\rm T}_\lambda$. In this case the polylogarithm in Eq. (\ref{c4}) has decomposition as 
\begin{equation}
\zeta_s(e^{-\beta |E_0|})=e^{-\beta |E_0|}+2^{-s}e^{-2\beta |E_0|}+...~,
\label{c6}
\end{equation}
where the first term leads to high accuracy for the function $\zeta_s(z)$. 
Thus the Eqs. (\ref{c4}) and (\ref{c5}) can be reduced to Eqs. (\ref{55}) and (\ref{57}) as 
\begin{equation}
\tilde{n}_{1}=\frac{1}{\lambda_q^{3}}\exp(-\beta |E_0|),~~~\tilde{n}_{0}=\tilde{n}_{q}-\tilde{n}_{1},
\label{c7}
\end{equation}
where  $\tilde{n}_{1}={\rm Lim}(\tilde{{\rm N}}_1/{\rm V})$  and $\tilde{n}_{0}={\rm Lim}(\tilde{{\rm N}}_0/{\rm V})$. 

The average energy of DQ's follows by Eqs. (\ref{40}) and (\ref{41}) as
\begin{equation}
{\rm U}_q=\langle\hat{{\rm H}}_q\rangle=-\left(\frac{\partial}{\partial\beta}{\rm ln}~\Xi_q\right)_{\mu_k=\tilde{\mu}}+\tilde{\mu} \tilde{{\rm N}}_0+\tilde{\mu} \tilde{{\rm N}}_1.
\label{c8}
\end{equation}

The entropy of diatomic quasiparticles in liquid ${\rm ^4He}$ is defined as ${\rm S}_q=-k_B\{{\rm Tr}(\hat{\rho}_q{\rm ln}\hat{\rho}_q)\}_{\mu_k=\tilde{\mu}}$. It can be reduced by Eqs. (\ref{40}) and (\ref{41}) to the form,

\begin{equation}
{\rm S}_q=k_B\left({\rm ln}~ \Xi_q-\beta\frac{\partial}{\partial\beta}{\rm ln}~ \Xi_q\right)_{\mu_k=\tilde{\mu}},
\label{c9}
\end{equation}
where the partition function $\Xi_q$ is given in Eq. (\ref{c2}).

\end{document}